\documentclass[useAMS,usenatbib]{mn2e}
\usepackage{graphicx}
\usepackage{multirow}
\usepackage{txfonts}
\usepackage{natbib}
\title[Dynamics of Swift~J1357.2-0933]
{Implications on accretion flow dynamics from spectral study of Swift~J1357.2-0933}
\author[S. Mondal and S. K. Chakrabarti]
{Santanu Mondal\thanks{santanu.mondal@uv.cl}$^{1,2}$ and Sandip K. Chakrabarti\thanks{chakraba@bose.res.in}$^{2,3}$\\ 
$^{1}$Instituto de F\'isica y Astronom\'ia, Facultad de Ciencias, Universidad de Valpara\'iso, Gran Bretana N 1111, Playa Ancha, Valpara\'iso, Chile\\
$^{2}$Indian Centre for Space Physics, Chalantika 43, Garia Station Rd., 
	     Kolkata, 700084, India\\ 
$^{3}$S. N. Bose National Centre for Basic Sciences, Kolkata, 700098, India} 

\begin{document}

\date{}

\maketitle

\begin {abstract}
We report a detailed spectral study of Swift~J1357.2-0933 low-mass X-ray binary during its 2017 outburst using {\it Swift} 
and {\it NuSTAR} observations. We fit the data with two component advective flow (TCAF) model and power-law model. 
We observe that the source is in hard state during the outburst, where the size of the 
Compton cloud changes significantly with disc accretion rate. 
The typical disc accretion rate for this source is $\sim 1.5-2.0~\%$ of the Eddington accretion rate $(\dot M_{Edd})$. 
The model fitted intermediate shock compression ratio gives an indication of the presence of jet, which is reported
in the literature in different energy bands. We also split NuSTAR data into three equal segments and fit with the model. 
We check spectral stability using color-color diagram and accretion rate ratio (ARR) vs. intensity diagram using different
segments of the light curve but
do not find any significant variation in the hardness ratio or in the accretion rate ratio. 
To estimate the mass of the candidate, we use an important characteristics of TCAF that the 
the model normalization always remains a constant. We found that 
the mass comes out to be in the range of $4.0-6.8~M_\odot$. From the model fitted results, we study the disc geometry 
and different physical parameters of the flow in each observation. 
The count rate of the source appears to decay in a time scale of $\sim 45 day$.
\end{abstract}



\section{Introduction} 

In a black hole binary, matter falls onto the compact object while transporting angular momentum outwards and 
mass inwards converting half of its gravitational potential energy into thermal energy and radiation energy.
Thus, it is interesting to study the accretion dynamics both in temporal and spatial domains. 
Unlike a persistent source, where accretion rates could be steady for a long
time, in an outburst source, rates are supposed to be varying.
Outbursting black hole candidates (BHCs) spend most of the times in the quiescence state 
and occasionally undergo bright X-ray outbursts, which are definitely due to a huge increase of accretion rate. 
One of the most natural ways to achieve this is by varying viscosity. 
It is possible that an outburst may be triggered by enhancement of viscosity at the
piling radius of matter. Chakrabarti (1990, 1996) suggested that when the rise of viscous process increases the
viscosity parameter above a critical value, the low angular momentum flow can acquire a Keplerian distribution
which becomes a SS73-like standard disk in presence of cooling (Shakura \& Sunyaev, 1973). Based on this
Chakrabarti (1997) concluded that the regular rise and fall of the accretion rate is due to the enhancement 
and reduction of viscosity at the outer regions of the disc.

Transient BHCs show several components in their spectra, namely, blackbody, power-law and an iron line at 
around 6.4\,keV. These distinct components are primarily from the optically thick and thin flow components.
It is observed that the spectra change their shape during the outburst following a cycle from hard to soft
through intermediate states i.e., see, the hardness intensity diagram (HID, Homan \& Belloni 2005; Nandi et al. 2012). 
Several attempts were made to explain changes in the spectral shape and its variation
(Remillard \& McClintock 2006 for a review). However, there was a lack of physical 
understanding behind this. Most of the studies are qualitative and phenomenological. It is believed
that changes in the accretion rate may be responsible for changes in spectral states (Maccarone \& Coppi 2003; 
Meyer-Hofmeister et al. 2004 and references therein). However, causes of the change in accretion rate on a 
daily basis, the origin of corona and its temperature, optical depth etc. were not very clear.

The problems were satisfactorily resolved when proper usage of the solution of transonic flows
in presence of viscosity was used. In a Two Component Advective Flow (TCAF, Chakrabarti \& Titarchuk 1995, hereafter, CT95) 
solution, a Keplerian disc which arises out of higher viscosity is immersed inside a hot sub-Keplerian flow of lower viscosity. 
This sub-Keplerian, low angular momentum hot matter forms an axisymmetric shock due to the dominance of 
the centrifugal force (Chakrabarti 1990; Molteni, Lanzafame \& Chakrabarti 1994).
The subsonic region between the shock boundary to the inner sonic point is hot and puffed up and behaves as the Compton cloud, 
which upscatters intercepted soft photons from the standard disc. This region also supplies matter to the jet and outflow 
(Chakrabarti 1999a, hereafter C99a). This is called the CENtrifugal barrier dominated BOundary Layer or CENBOL.
This region could be oscillatory when its cooling time scale roughly matches with the 
infall (compression) timescale inside CENBOL  (MSC96; Chakrabarti \& Manickam 2000; Chakrabarti et al. 2015). 
Recently, Chakrabarti et al. (2015) applied the resonance condition for H~1743-322 black hole candidate and 
showed that the low frequency quasi-periodic oscillations (LFQPOs)
are produced when cooling time scale roughly matches with the infall time scale. 
Transonic solution by Mondal \& Chakrabarti (2013; CT95)
shows that cooling mechanism is also responsible for the change in spectral states. 
The presence of two components as in CT95 is established
by many other authors in the literature (Smith et al. 2001, 2002; Wu et al. 2002; Ghosh \& Chakrabarti, 2018).

After the implementation of TCAF in XSPEC (Debnath et al. 2014) and fitting data of several black hole candidates, 
one obtains physical parameters of the underlying inflow, such as the accretion rates of the disk and halo components, 
the shock location and the shock strength. If the mass is unknown, this will also be found out from the spectral fit
(e.g., Molla et al. 2016; Bhatterjee et al. 2017).
A plot of photon count variation with accretion rate ratio (ARR) gives the so call ARR intensity diagram 
(ARRID, Mondal et al. 2014b; Jana et al. 2016) and directly shows why the spectral state changes.
The changes in accretion rates on a daily basis is due to changes in viscosity parameter during the outburst (Mondal et al. 2017). 
The time scale of the changes can also be estimated from the model fitted accretion rates (Jana et al. 2016). 
From the spectral fits, QPO frequencies can be predicted as well (Debnath et al. 2014; Chakrabarti et al. 2015;
Chatterjee et al. 2016).

Till date, many faint X-ray binaries have been discovered and with even fainter companions.
Swift~J1357.2-0933 has one of the shortest orbital periods and is a very faint black hole X-ray transient. The source was 
first detected in 2011 by the {\it Swift} Burst Alert Telescope (Barthelmy et al. 2005; Krimm et al. 2011). The distance to the 
source is not confirmed. This can range from $\sim 1.5$ - 6.3~kpc (Rau et al. 2011; Shahbaz et al. 2013). 
There is also a large discrepancy in mass measurement of the source. 
The mass of the black hole is estimated to be $> 3.6~M_\odot$ by 
Corral-Santana et al. (2013) and  $> 9.3~M_\odot$ by Mata S\`anchez et al. (2015). Corral-Santana et al. (2013) also estimated
the orbital period to be $2.8\pm0.3~hrs$ from the time-resolved optical spectroscopy. They observed recurring dips on 2-8 min 
time-scales in the optical lightcurve, although the RXTE and XMM-Newton data do not show any of the above evidences 
(Armas Padilla et al. 2014). The observed broad, double-peaked $H_{\alpha}$ profile supports a high orbital inclination 
(Torres et al. 2015). Very recently, Russell et al. (2018) found an evolving jet synchrotron emission using long term 
optical monitoring of the source.

In the earlier outburst during 2011, the source had a variable accretion and showed very regular temporal and spectral evolution. 
The detailed multiwavelengh lightcurve is studied by Weng \& Zhang (2015). Recently, Swift~J1357.2-0933 showed 
renewed activity on 2017 April 20 (Drake  et al. 2017) and April 21 (Sivakoff et al. 2017, observed by Swift/XRT). 
Very recently, Stiele \& Kong (2018) observed the source by NuSTAR and there is a simultaneous observation in 
Swift/XRT as well. Thus the present outburst covers a broadband energy range. In this paper, we use the above data  
to study the flow dynamics of the source during its 2017 outburst.

The paper is organized as follows: in the next Section, we present the observation and data analysis procedure. 
In \S 3, we discuss about the model fitted results and geometry of the disc during the outburst. 
We also estimate the mass from the model fit. In \S 4, we calculate various
physical quantities of the disc from model fitted parameters to infer about the disc properties. 
Finally, in \S 5, we draw our brief concluding remarks.
  
\section{Observation and data analysis}

In the present manuscript, we analyze both Swift/XRT (Gehrel et al., 2004) and NuSTAR 
(Harrison et al. 2013) satellite observations of the BHC Swift~J1357.2-0933 during its 2017 outburst. 

\subsection{Swift} 
In our present analysis, we use 0.5-7.0~keV Swift/XRT observation of Swift~J1357.2-0933 
during 2017 outburst, the timings of which overlap with the NuSTAR observations presented below.
The observation IDs are 00088094002 (Photon Counting mode, PC) and 00031918066 (Windowed Timing mode, WT). 
We use {\it xrtpipeline v0.13.2} task to extract the event fits file from the raw XRT data. All filtering tasks 
are done by {\it FTOOLS}. To generate source and background spectra we run {\it xselect} task. We use
swxpc0to4s6\_20130101v014.rmf and swxwt0to2s6\_20130101v015.rmf file for the response matrix. 
The $\it grppha$ task is used to group the data. We use 10 bins in each group. We use same binning for both the observations
and also in NuSTAR data, which we discuss in the next section.

\subsection{NuSTAR}
We analyze two NuSTAR observations with observation IDs: 90201057002 (hereafter O1, combined with 00088094002) and 
90301005002 (hereafter O2, combined with 00031918066) of BHC Swift~J1357.2-0933 with 
energy range 2.0-70~keV. NuSTAR data were extracted using the standard NUSTARDAS v1.3.1 software. 
We run `nupipeline' task to produce cleaned event lists and `nuproducts' for spectral file generation. 
We use $30^{''}$ radius region for the source extraction and $45^{''}$ for the background
using ``ds9''. The data is grouped by ``grppha'' command, where we group the whole energy bin with 10
bins in each group. We choose the same binning and fitting criteria for both the observations. 
However, the data quality of O2 is not good and above $\sim 20$~keV it is highly noisy.
Thus the count at different energy ranges in O2 is not similar to O1. We split the NuSTAR
observations into three different time ranges (S1, S2, and S3) each of which contains $\sim 24~ksec$ (for O1) 
and $\sim 15~ksec$ (for O2) data. For that purpose, first we make our own ``GTI'' files for each time range 
using ``gtibuild'' task in SAS environment and use those GTI files during data extraction.
For spectral fitting of the data we use XSPEC (Arnaud, 1996) version 12.8.1. 
We fit the data using 1) Power-law (PL), and 2) TCAF models. The detailed spectral fitting with other 
phenomenological models and with reflection model (relxill) are discussed in Stiele \& Kong (2018). 
Using relxill model and assuming high inclination, they found that the disc is truncated close to the 
black hole independent of spin parameter. Here, we mainly focus on the TCAF model fitted parameters to study 
the physical properties of the disc and its geometry during the outburst. 
To fit the spectra with the TCAF model in XSPEC, we have a TCAF model generated {\it fits} file (Debnath et al. 2014).
We follow the same analysis procedure for TCAF fitting with 
Swift and NuSTAR data as discussed in Mondal et al. (2016). We use the absorption model $\it{TBabs}$ (Wilms et al. 2000) 
with hydrogen column density fixed at 1.3$\times$~10$^{21}$~atoms~cm$^{-2}$ throughout the 
analysis. In the above absorption model solar abundance vector set to ``wilm'', which includes cosmic 
absorption with grains and $H_2$ and those absorptions which are not included in the paper are set to zero. 
We keep the mass as a free parameter in order to estimate it from the TCAF itself.
  
\section{Model fitted results and disc geometry}

We study the BHC Swift~J1357.2-0933 using Swift/XRT and NuSTAR observations with PL model. 
PL model fitted photon index is $\sim 1.7-1.8$. Thus the object is in a hard state.
We also fit the data with the TCAF solution based fits file which uses five physical parameters:
The parameters are as follows (i) Mass of the black hole,
(ii) disc accretion rate, (iii) halo accretion rate, (iv) location of the shock, and (v) shock
compression ratio. Parameters (ii) to (v) collectively give the electron density and temperature, photon spectrum and 
density, the fraction of photons intercepted by the CENBOL from the disk, as well as the reflection 
of hard photons from the CENBOL by the disc. All of these depend on the mass 
of the black hole. According to CT95, if one increases the halo rate keeping other parameters frozen, 
the model will produce a hard spectrum. Similarly, increasing the disc rate 
leaving other parameters frozen, will produce a soft spectrum. When the location of the 
shock is increased keeping other parameters fixed, spectrum will be harder.
A similar effect is seen for compression ratio also (see also, Chakrabarti 1997). 
In general, in an outburst, all the parameters will change smoothly in a multidimensional space. 
The model fitted results for both the observations are given in Table~1. In Fig.~1(a-b), we present the TCAF model fitted spectra. 
The model fitted parameters show that the disc rate was higher on the second day. Opposite is true for the halo rate. In both the 
observations, the ratio of the halo rate and disc rate is larger than unity, i.e., the flow is dominated by the 
sub-Keplerian rate. This is an indication of the hard state. At some point in time, before the 
second observation day, viscosity may have started to go up, and the Keplerian disc rate also started to increase.
However, this was not enough so as to change the spectral state (as a minimum viscosity is required for such changes
(Mondal et al. 2017). The PL model fitted photon index also indicates a hard spectral state.

As the shock is the outer boundary of the Compton cloud, the movement of the shock shows a change in size of the 
Compton cloud. On the second day of observation, the shock moved closer to the black hole as compared to the first day from
$X_s=81.00~r_g$ to $\sim 36.66~r_g$ (where, Schwarzschild radius $r_g=2GM_{BH}/c^2$). This type of behavior is observed routinely in all
the black holes during the rising phase of the outburst. The behavior of advancing inner edge of the disc was 
studied by several authors for various outbursting candidates 
(Esin et al. 1997; Tomsick et al. 2009; Dutta \& Chakrabarti 2010; Shidatsu et al. 2011; Nandi et al. 2012).
The shock compression ratio (the ratio of the post-shock to pre-shock flow density experienced by the low angular 
momentum component) is always observed to be higher than unity and is generally of intermediate strength. In this
case, the jets and outflows are expected to be strong (C99a; Chakrabarti 1999b). For the compression
ratio given in the Table, the expected outflow/inflow ratio will be $3.4-4.2~\%$. 
This jet may appear and disappear during the transition phase of the outburst, although the actual 
interrelation between the jet properties and the X-ray spectral state evolution is still debate (Kalemci et al. 2005; Dincer et al. 2014).
Several attempts have been taken (Tomsick et al. 2009; Petrucci et al. 2014 and references therein) to understand the evolution of the 
Compton cloud and spectral state transitions as a function of luminosity during the outburst. Recent transonic 
solution of Mondal et al. (2014a) following the same flow geometry and jet configuration of C99a, showed that since the
jet removes a significant amount of matter from CENBOL, it is easier to cool the Compton cloud. Thus, a change in
spectral states in presence of jet is expected to be faster. Following the above model understanding and the values of 
the model fitted parameters, we conclude that the source was in rising hard state of the outburst during the observations.

\begin{figure}
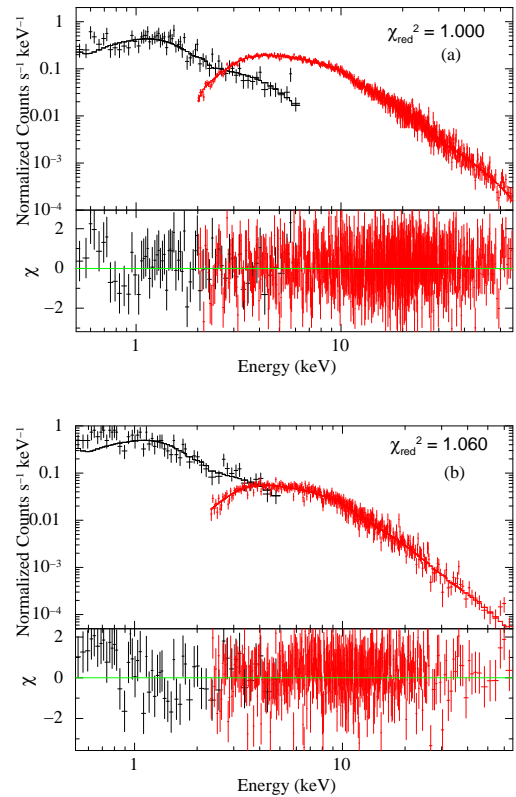

\centering{
\includegraphics[height=7.0truecm,angle=270]{fig-data1-rev2.ps}
\includegraphics[height=7.0truecm,angle=270]{data2-tcaf.ps}}
\caption{Swift/XRT and NuSTAR (focal plane module A, FPMA) data from 0.5-70.0~keV energy range are 
fitted with an absorbed TCAF model. The model fitted parameters are given in Table~1. } 
\label{fig1ab}
\end{figure}

\begin{table}
\vskip -0.4 cm
\centering
\caption{\label{table1} PL and TCAF model fitted parameters of combined Swift/XRT and NuSTAR observation}
\begin{tabular}{|l|l|l|cc|}
\hline
Obs.          & model1 (PL)              & model2 (TCAF)      \\
\hline
\hline
1             &$\Gamma=1.678\pm0.007$     &$M_{BH}=6.84\pm0.39$       \\
              &$norm=0.018\pm0.001$       &$\dot{m}_d=0.017\pm0.001$   \\
              &$\chi^2/dof=792.95/798$   &$\dot{m}_h=0.378\pm 0.021$   \\
              &--                         &$X_s=81.04\pm7.91$          \\
              &--                         &$R=2.74\pm0.34$               \\
              &--                         &$\chi^2/dof=804.56/805$      \\
\hline
2             &$\Gamma=1.811\pm0.015$     &$M_{BH}=4.01\pm0.34$       \\
              &$norm=0.0088\pm0.0003$     &$\dot{m}_d=0.019\pm0.004$   \\
              &$\chi^2/dof=451.98/442$    &$\dot{m}_h=0.213\pm 0.013$   \\
              &--                         &$X_s=36.66\pm4.42$          \\
              &--                         &$R=2.0\pm0.16$               \\
              &--                         &$\chi^2/dof=473.12/446$      \\
\hline
\end{tabular}
\leftline{$M_{BH}$ is in unit of $M_{\odot}$ and Mass accretion rates are in $\dot{M}_{Edd}$ unit.}
\leftline{$X_s$ is in $2GM_{BH}/c^2$ unit.}
\end{table}

In Fig.~2, we show the hardness intensity diagram (HID) and accretion rate ratio intensity diagram (ARRID) 
after splitting the data in three segments of equal time interval. The HID shows that
the hardness ratio varied from 1.7 to 2.2. Thus one can say 
that the source is moderately variable. After fitting the data segments with TCAF model, we see
that the ARR in ARRID varies by a factor of about 2. 
From both the diagrams, we note that the flow was not rapidly evolving.
\begin{figure}
\centering{
\includegraphics[height=7.truecm,angle=0.0]{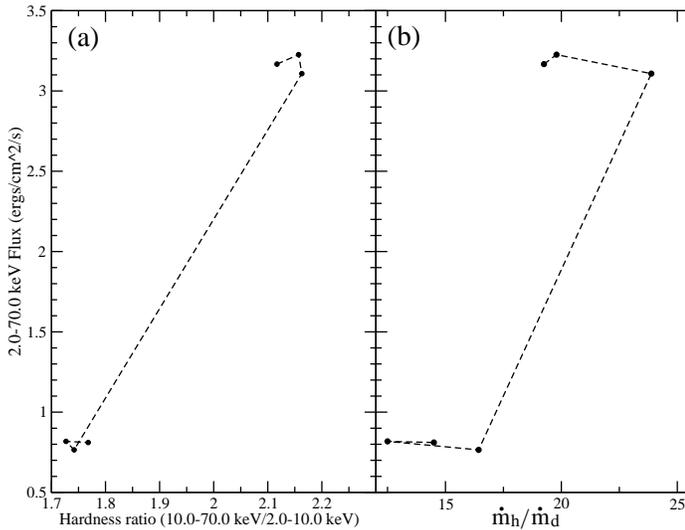}}
\caption{(a) Hardness ratio (10.0-70.0~keV/2.0-10.0~keV) of NuSTAR data for both the observations after splitting 
the data in three intervals mentioned in Sec.~2. Total flux is in 1.0E-10~$ergs/cm^2/s$ unit. 
	(b) Variation of total flux with ARR. It is to be noted
that the hardness ratio varies moderately between 1.7-2.2 and ARR by a factor of 2.
} 
\label{fig2ab}
\end{figure}

In a series of papers with TCAF model fits, constant normalization was used for a given object observed 
by a given instrument as it is a conversion factor between the observed flux and the model flux. 
In Fig.~1, to fit the data we obtained model normalization $1.12$ for NuSTAR observations, whereas Swift observations 
the fitting procedure produces the normalization values of $0.31$ for O1 and $1.13$ for O2.
The freedom in absorption does not improve the fit significantly, rather gives a factor of $10$ difference 
in $N_H$ between O1 and O2. As the normalization is not chosen by hand and it comes out as a fitted parameter, 
the ratio of the normalizations is important.
The relative normalization for the two instruments can be calculated from the ratio of the model normalizations obtained from the fits
of each instrument's spectrum. There could be several reasons for getting different normalizations in Swift data in O1 and O2: 
1) if the Swift and NuSTAR data are not strictly simultaneous,
2) there could be a pile-up issue, and 3) if the source is close to the PC/WT switch point, then it may be relatively faint for WT.
As the source was not highly variable during that period, the point~1 cannot be a reason for the difference in normalization.
The pile-up could be an issue as the count rate is above $0.5$ cnts/sec. We should mention that we have performed the
analysis using pile-up corrected spectrum files for PC mode generated by online Swift/XRT product generator (Evans et al. 2009).
However, the difference in normalization persists. The model fitted parameter values also remained unchanged with
an acceptable reduced $\chi^2 (\sim 1.0)$ keeping the mass of the source
within $4.0~M_\odot - 6.8~M_\odot$. The third point should not be the reason for difference in normalization
as the source is sufficiently bright for WT mode.

These leads us to conclude that the difference in normalization could be due to other physical processes 
inside the accretion disc which were not incorporated in TCAF. 
As discussed in Jana et al. (2016) and Molla et al. (2017), the presence of jet is a reason
for the difference in normalization. During the present outburst period, 
activity in jet was observed for this source. Thus the presence of jet/outflow
could be a reason behind the variation of normalization constant.

\section{Estimation of different physical quantities of the disc} 

In this Section, we estimate some physical parameters of the disc using TCAF model fitted parameters.
From Kepler's law, one can derive a relation between orbital period ($P$) and orbital separation $a$ as follows:
$$
a=3.5\times10^{10}m_2^{1/3}(1+q)^{1/3}P(hr)^{2/3},
\eqno{(1)}
$$
where $q$ ($=m_1/m_2$) is the mass ratio of the component stars. Outer disc radius ($r_{out}$) of the
primary star is calculated from the Roche lobe radius of the primary following Eggleton (1983) as,
$$
r_{Rl,1}=\frac{0.49\times~a~q^{2/3}}{0.6\times~q^{2/3}+ln[1+q^{1/3}]}.
\eqno{(2)}
$$
In this work, we consider that the outer edge of the disc ($r_{out}$) is 70\% of the Roche lobe radius.
We use the values of P (= 2.8~hrs) and companion mass ($= 0.17~M_\odot$) in Eq. ~2, already derived by 
Corral-Santana et al. (2013), to estimate Roche lobe radius, which appears to be $r_{out}=0.68\times10^{11}$. 
Using the above derived outer disc radius and model fitted accretion rate, we  
calculate kinematic viscosity ($\nu$) and surface density ($\Sigma$) of the disc. It is to be noted that 
the disc accretion rate which we are using from TCAF fit is constant throughout the disc, 
thus we assume that the accretion at $r_{out}$ is same as the model fit value. We use standard disc equations 
(SS73) below to derive the above two physical parameters ($\nu$ and $\Sigma$):
$$
\Sigma\simeq \frac{\dot{M}_{d}}{3\pi\nu},
\eqno{(3)}
$$
where, $\nu$ is calculated using $SS73\alpha$ disc model ($\nu$=$\alpha C_s H$). 
Here $C_s^2(=\frac{kT}{\mu m_p})$ and $H (=\frac{C_s}{\Omega_K} = 0.024)$ are the sound speed at 
the outer radius of the disc, height of the disc and $\Omega_K$ is the Keplerian angular velocity at 
that point respectively. The temperature of the disc is calculated from the disc accretion rate. 
The estimated values of $\Sigma$ and $\nu$ are $\sim 49.1~gm/cm^2$ 
and $\sim 4.0\times10^{14}~cm^2/sec$ respectively. During our calculation, we consider that the disc is not 
self-gravitating i.e., vertical hydrostatic equilibrium is maintained against the pull of the gravity. 
The computed low surface density and kinematic viscosity are indicating a stable disc. 
In the context of disc stability, Weng \& Zhang (2015) mentioned that the high ratio of near UV luminosity to X-ray 
luminosity indicates that the irradiation is unimportant in this outburst, while the near-exponential decay profile and the long 
decay time-scale conflicts with the disc thermal-viscous instability model. Hence they suggested that the disc 
is thermally stable during the outburst. Armas Padilla et al. (2013) also found that the correlation between Swift/UVOT 
v-band and XRT data is consistent with a non-irradiated accretion disc. 

Here, we present the chain of logical steps used to fit the observed lightcurve: (i) we have mass and disc 
accretion rate from TCAF fit, (ii) using (i) we estimate disc temperature thus the sound speed, (iii) we also have orbital period ($P$) 
and mass function ($q$) from literature, (iv) outer disc radius is estimated using Eq.~1 from the parameters 
of (iii), (v) once (ii) and (iv) are known one can estimate height of the disc, Keplerian 
angular frequency, and disc kinematic viscosity to estimate the viscous time scale, and 
(vi) finally, we extract SS73-$\alpha$ parameter for which the derived and the fitted $\tau$ values are consistent. 
Here, $\alpha$ takes the value 0.25 to give a decay timescale ($\tau$) of $\sim 45$~days.
To fit the observed count rate using decay timescale, we use an exponential decay function, which is given by:
$$
f=A~exp(-t/\tau),
\eqno{(4)}
$$
where $A$ is a normalization constant, which takes the value of $3.91\pm0.16$ with exponential decay timescale ($\tau$) around 
$45.28\pm4.78~days$. The estimated SS73 disc viscosity parameter ($\alpha=0.25$) becomes same order as those obtained 
for other observed candidates (Nagarkoti \& Chakrabarti 2016; Mondal et al. 2017). Here we consider $A$ as a constant, however it should depend on the source distance and the physical properties of the disc. The detailed calculation is beyond the scope of this paper. In Fig.~3, we show exponential decay function fitted with the observed count rate. 

\begin{figure} 
\centering{
\includegraphics[height=7.truecm,angle=0]{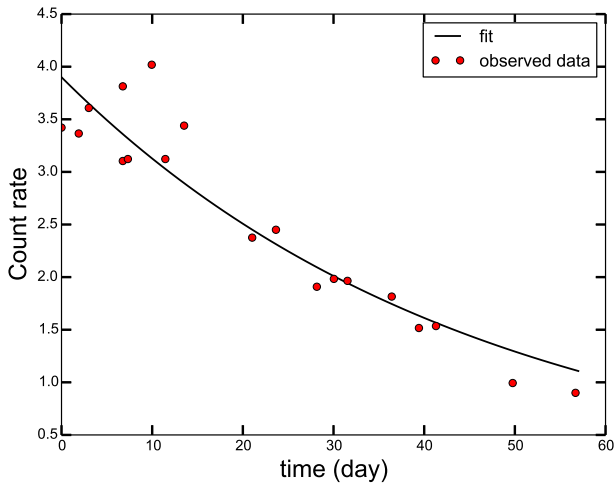}}
\caption{Variation of count rate with time (day). Black solid line shows the model fit and filled circles are observed 
count rates. Data are fitted with $H/R=0.024$ (from above), $P=2.8$~hrs, $q=0.03$ and SS73 $\alpha=0.25$. Count rate of Swift lightcurve is adopted from Stiele \& Kong (2018). 
} 
\label{fig3a}
\end{figure} 

\section{Summary and Conclusions}

In this paper, we analyzed Swift/XRT and NuSTAR spectra of a known Galactic stellar mass black hole source
Swift~J1357.2-0933 during its 2017 outburst using a phenomenological PL model and a physical TCAF solution. 
We find that on both the observation dates, the sub-Keplerian halo accretion rate is higher than the Keplerian accretion rate. 
In the second observed day, the disc rate is increased as compared to first observed day and 
opposite variation is seen in the sub-Keplerian rate. The object was in hard state on both the days.
As the halo rate is higher than the disc rate and the shock compression ratio 
is always greater than unity, the shock moved inward due to cooling. This could be the signature of the hard state 
in the rising phase of the outburst. The shock was seen to move at $\sim 0.15$~m/s which is similar 
to the shock velocity in other outbursts (Debnath et al. 2010 for GX~339-4; Mondal et al. 2015 for H~1743-322 and 
references therein). This indicates that the outburst probably remained in the rising phase and 
no other spectral state has been missed in between these $\sim 40~days$. 
It is to be noted that the companion of this candidate is a star evolved through nuclear fusion
(Shahbaz et al. 2013) with an initial mass $\sim 1.5~M_\odot$, which has evolved to $0.17~M_\odot$. 
Thus there is a possibility that the accretion is mostly 
dominated by companion winds and thus the halo rate is always higher. Our model fitted 
disc rate is $\sim 1.5-2$\% of $\dot M_{Edd}$. As the disc rate increases and the shock location decreased in 
$\sim 40$~days, viscosity must have gone up since the first observation. As the shock compression ratio is in intermediate 
strength, in this case, the jets and outflows are expected to be strong with outflow/inflow rate ratio 3.4-4.2\%.
From TCAF model fit, we estimate the mass range for this black hole 
candidate to be $4.0-6.8~M_\odot$. However, a few more observations would have reduced the 
error-bar significantly.

We also study different physical parameters of the disc. For that, we calculate the surface density, kinematic viscosity and
disc aspect ratio etc. of the disc using the model fitted parameters. We find that the disc surface density is not high enough 
signifying that the disc is stable in nature. The estimated surface density is also reasonable to produce a 
consistent $\alpha$ value to study the decay of the lightcurve. 
We find that the lightcurve fits with exponential decay function with the decay time scale of $\sim 45~day$, which is consistent with
the derived decay time scale when $\alpha$=0.25. 
Thus from the model fit we can study the spectra, disc properties, lightcurve decay and estimate viscosity parameter at a time. 
  
\section{Acknowledgment}
We thank anonymous referee for useful comments on the manuscript. 
SM acknowledges Swift team (especially Kim Page) for useful discussions on Swift data fitting and Patricia Ar\`evalo
for commenting on the preliminary version of the paper.
SM acknowledges FONDECYT fellowship grand (\# 3160350) for this work.
This research has made use of the NuSTAR Data Analysis Software (NuSTARDAS) jointly developed by the ASI Science
Data Center (ASDC, Italy) and the California Institute of Technology (Caltech, USA),
and the XRT Data Analysis Software (XRTDAS) developed under the responsibility of the ASI Science 
Data Center (ASDC), Italy. This research has made use of data obtained through the High
Energy Astrophysics Science Archive Research Center Online Service, provided by the NASA/Goddard Space Flight Center.
{}
\end{document}